\documentclass{jfm}

\usepackage{graphicx}
\usepackage{epstopdf}
\usepackage{natbib}
\usepackage{amsmath}
\usepackage{amssymb}
\usepackage{mathtools}
\usepackage{bm}
\usepackage{mathrsfs}
\usepackage{colortbl}
\usepackage[svgnames,x11names]{xcolor}
\usepackage[font=small,labelfont=bf]{caption}
\usepackage{multirow}
\usepackage{booktabs}
\usepackage{bibentry}
\nobibliography*

\ifCUPmtlplainloaded \else
  \checkfont{eurm10}
  \iffontfound
    \IfFileExists{upmath.sty}
      {\typeout{^^JFound AMS Euler Roman fonts on the system,
                   using the 'upmath' package.^^J}%
       \usepackage{upmath}}
      {\typeout{^^JFound AMS Euler Roman fonts on the system, but you
                   dont seem to have the}%
       \typeout{'upmath' package installed. JFM.cls can take advantage
                 of these fonts,^^Jif you use 'upmath' package.^^J}%
      }
  \else
  \fi
\fi

\ifCUPmtlplainloaded \else
  \checkfont{msam10}
  \iffontfound
    \IfFileExists{amssymb.sty}
      {\typeout{^^JFound AMS Symbol fonts on the system, using the
                'amssymb' package.^^J}%
       \usepackage{amssymb}%
       \let\le=\leqslant  
       \let\ge=\geqslant  
      }{}
  \fi
\fi

\ifCUPmtlplainloaded \else
  \IfFileExists{amsbsy.sty}
    {\typeout{^^JFound the 'amsbsy' package on the system, using it.^^J}%
     \usepackage{amsbsy}}
    {}
\fi

\newcommand\Rey{\mbox{\textit{Re}}}  

\newsavebox{\astrutbox}
\sbox{\astrutbox}{\rule[-5pt]{0pt}{20pt}}

\newcommand{\vol}{\mathop{\ooalign{\hfil$V$\hfil\cr\kern0.08em--\hfil\cr}}\nolimits}

\newcommand\Wo{\mathit{Wo}}
\newcommand\A{\mathit{A}}

\title[The effect of pulsation frequency on transition in pulsatile pipe flow]{The effect of pulsation frequency on transition in pulsatile pipe flow}

\author[Duo Xu and Marc Avila]%
{Duo Xu$^{1,2}$\thanks{Email address for correspondence: duo.xu@zarm.uni-bremen.de},\ns
and
Marc Avila$^{1,2}$}

\affiliation{
$^1$ University of Bremen, Center of Applied Space Technology and Microgravity  (ZARM), 28359 Bremen, Germany\\[\affilskip]
$^2$ Friedrich-Alexander-Universit\"{a}t Erlangen-N\"{u}rnberg, Institute of Fluid Mechanics (LSTM), 91058 Erlangen, Germany\\[\affilskip]
}

\pubyear{?}
\volume{?}
\pagerange{?}
\date{?; revised ?; accepted ?. - To be entered by editorial office}
\begin{document}

\maketitle

\begin{abstract}
Pulsatile flows are common in nature and in applications, but their stability and transition to turbulence are still poorly understood. Even in the simple case of pipe flow subject to harmonic pulsation, there is no consensus among experimental studies on whether pulsation delays or enhances transition. We here report direct numerical simulations of pulsatile pipe flow at low pulsation amplitude $\A\le 0.4$. We use a spatially localized impulsive disturbance to generate a single turbulent puff and track its dynamics as it travels downstream. The computed relaminarization statistics are in quantitative agreement with the experiments of \citeauthor{Xu17a} (\textit{J. Fluid Mech.}, vol. 831, 2017, pp. 418--432) and support the conclusion that increasing the pulsation amplitude and lowering the frequency enhance the stability of the flow. In the high-frequency regime, the behaviour of steady pipe flow is recovered. In addition, we show that when the pipe length does not permit the observation of a full cycle, a reduction of the transition threshold is observed. We obtain an equation quantifying this effect and compare it favourably with the measurements of \citeauthor{Stettler86} (\textit{J. Fluid Mech.}, vol. 170, 1986, pp. 169--197). Our results resolve previous discrepancies, which are due to different pipe lengths, perturbation methods and criteria chosen to quantify transition in experiments. 
\end{abstract}


\section{Introduction}\label{sec:introduction}

\begin{figure}
\centering
\includegraphics[width=0.8\textwidth, trim={1.1cm 1cm 0.7cm 0cm}, clip]{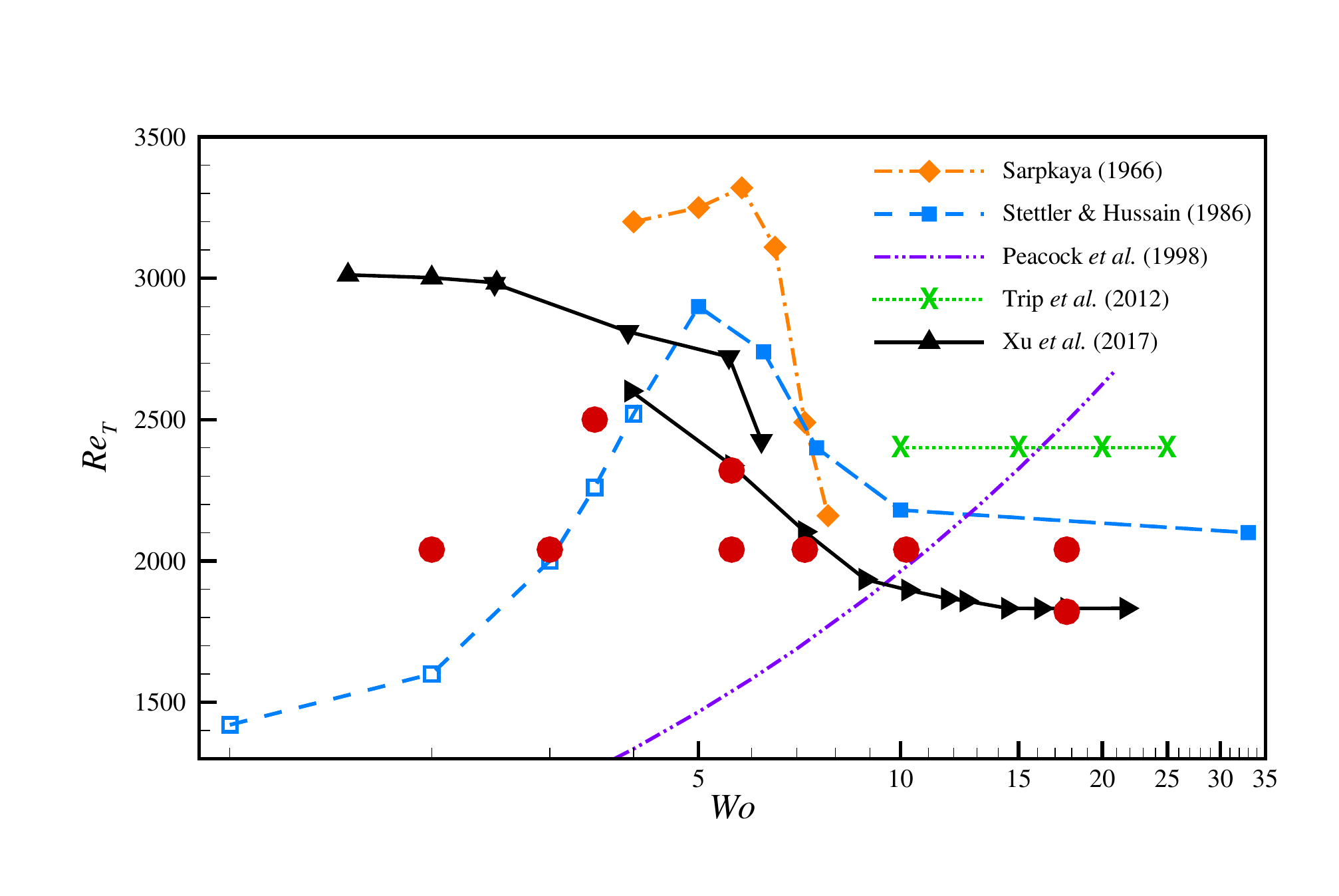}
\caption{\label{fig:exp} (Colour online) Transition thresholds from experiments of pulsatile pipe flow with pulsation amplitude  $A=0.4$. Hollow symbols denote results obtained with a measurement length preventing the observation of a full cycle before turbulence reaches the measurement point (see text and table~\ref{tab:exp}). Lines connect experimentally measured transition points and are to guide the eyes. The data of \citet{Xu17a} were obtained with three different pipe lengths, and the data from each pipe length are connected separately and shown in different triangle styles. \citet{Peacock98} proposed an equation for the transition threshold, $\Rey_T=1220\Wo^{0.42}A^{0.17}/(1+A)$, which is shown here for $\A=0.4$. \cite{Trip12} noted that the transition threshold was independent of the pulsation amplitude and frequency and remained identical to that of steady pipe flow. Details of the experimental set-ups are given in table~\ref{tab:exp}. The filled (red) circles depict the parameter groups studied in this work.
}
\end{figure}

Pulsatile flow through conduits and tubes is found in the cardiovascular system and is commonplace in hydraulic systems because of the mechanisms pumping the fluid. Whether pulsatile flows are laminar or turbulent is relevant to physiology because the formation of aneurysms and other diseases are associated with the transition to turbulence \citep{Freis64,chiu2011}. We here focus on the simplest pulsatile flow: fluid is driven through a straight cylindrical pipe at an unsteady flow rate. Despite recent advances in understanding the transition to turbulence for steady pipe flow (see e.g.~\citet{Barkley16} for a recent review), the stability and transition of pulsatile pipe flow remain poorly understood to date. In the steady case, the only governing parameter is the Reynolds number $\Rey=U_sD/\nu$, where $U_s$ is the mean speed, $D$ the pipe diameter and $\nu$ the viscosity of the fluid, whereas in the pulsatile case the stability depends also on the pulsation amplitude $\A=U_o/U_s$, where $U_o$ is the oscillatory component of the speed, and on the Womersley number $\Wo=D/2\sqrt{\omega/\nu}$, where $\omega$ is the pulsation frequency. 

The transition thresholds determined in several experimental studies of pulsatile pipe flow with amplitude $\A=0.4$ are shown in figure~\ref{fig:exp}. \citet{Peacock98} performed measurements in wide ranges of parameter values and proposed a master equation for the transition threshold. Their equation suggests that the transition threshold $\Rey_T$ increases monotonically as the frequency is increased. However, \citet{Stettler86, Trip12, Xu17a} agree that, as the pulsation frequency becomes large, the dynamics and transition threshold from the steady case, $\Rey_{T,s}$, are recovered. In the intermediate- and low frequency regimes there are strong discrepancies between the different experimental studies. \citet{Sarpkaya66} found that $\Rey_T$ increases and then decreases as $\Wo$ is decreased, with a maximum of $\Rey_T$  around $\Wo\approx 5$. A similar behaviour was also reported by \citet{Stettler86}, who reduced $\Wo$ down to $1$.  By contrast, \citet{Xu17a} found that $\Rey_T$ increases rapidly and gradually approaches an upper limit as $\Wo$ is decreased. 

It is worth noting that all the aforementioned experiments were conducted at $\Rey$ far below the linear instability thresholds determined by \citet{Thomas11}, indicating that transition was triggered by finite-amplitude perturbations.  This is exactly the same situation as in steady Poiseuille flow. It implies that the transition threshold $\Rey_T$ depends on the type of perturbation employed \citep{Peixinho07}, or unknown source of noise or imperfections in the absence of an explicit disturbance. \citet{Peacock98} did not explicitly disturb the flow, whereas \citet{Stettler86, Trip12} used orifices, thereby perturbing the flow continuously. By contrast, \citet{Sarpkaya66} and  \citet{Xu17a} used impulsive perturbations. The former periodically  moved   a wire inserted in the pipe, whereas the latter used a fluid injection. These differences explain why, as $\Wo$ becomes large, the transition threshold saturates at different values in the experiments of \citet{Stettler86, Trip12, Xu17a}. An overview of the disturbances and other details of the aforementioned experiments is given in table~\ref{tab:exp}. 

\begin{table}
  \centering
  \begin{tabular}{@{\hspace{0pt}} p{60pt} @{\hspace{10pt}} p{20pt} p{20pt} @{\hspace{10pt}} p{55pt} p{50pt} @{\hspace{10pt}}  p{35pt} p{30pt} @{\hspace{25pt}} p{30pt} @{\hspace{-2pt}}}

    {Reference} & \multicolumn{2}{c}{Pipe} & \multicolumn{2}{c}{Measurement} & \multicolumn{2}{c}{Perturbation} & {$\Rey_{T,s}$}\\
    \cmidrule(r){2-3} \cmidrule(r){4-5} \cmidrule(r){6-7}
    & ${{L}}$ & ${{\Wo^*}}$ & {{Technique}} & {{Quantity}} & {{Device}} & {{Method}} & \\
    \addlinespace[2mm]
    \citet{Sarpkaya66} & $1020$ & $2.29$ & {\raggedright pressure\\ sensor} & {\raggedright length of\\ turbulence} & tipping wire & impulse & $2100$\\ 
    \citet{Stettler86} & $330$ & $4.02$ & LDV & velocity &orifice & continuous & $2100$\\ 
    \citet{Peacock98} & $100$ & $7.31$ & hot-film & shear stress & none & none & $2000$ \\ 
    \citet{Trip12}     & $150$ & $5.97$  & PIV & {\raggedright turbulence \\ intermittency} &orifice & continuous & $2400$\\ 
    \citet{Xu17a} & {\raggedright$2250$\\$1300$\\$350$} & $1.54$ & visualization & {\raggedright survival \\ probability\\ $P=0.5$} & injection & impulse & {\raggedright $1905$\\ $1890$ \\ $1855$} \\ 

  \end{tabular}
    \caption{\label{tab:exp} Experimental studies of pulsatile pipe flow. Here $L$ is the dimensionless distance (in diameters) between perturbation and  downstream measurement point; $\Wo^*$ is the smallest Womersley number that allows observation of a full pulsation cycle in a pipe for a given length $L$ (see text). LDV and PIV stand for laser Doppler velocimetry and particle image velocimetry, respectively; and $\Rey_{T,s}$ is the transition threshold for steady flow.}

\end{table}

As pointed out by \citet{Xu17a}, a crucial specification in pulsatile flow is the pipe length: the dynamics of turbulence changes over the pulsation cycle as turbulence travels down the pipe. Hence, in order to quantify the overall pulsation effect on transition, the distance between disturbance and measurement point must be long enough to let turbulence experience at least one full pulsation cycle. Assuming that turbulent structures travel at the mean speed $U_s$, the minimum feasible Womersley number  is $\Wo^*=\sqrt{\Rey\pi/(2 L)}$, where $L$ is the distance (in diameters). Table~\ref{tab:exp} gives the values of $\Wo^*$ of the aforementioned experiments, where $\Rey=3400$ was used to estimate $\Wo^*$. The hollow symbols in figure~\ref{fig:exp} denote experiments performed in insufficiently long pipes. 

\begin{figure}
\centering
\includegraphics[width=\textwidth, trim={0cm 0cm 0.4cm 0cm}, clip]{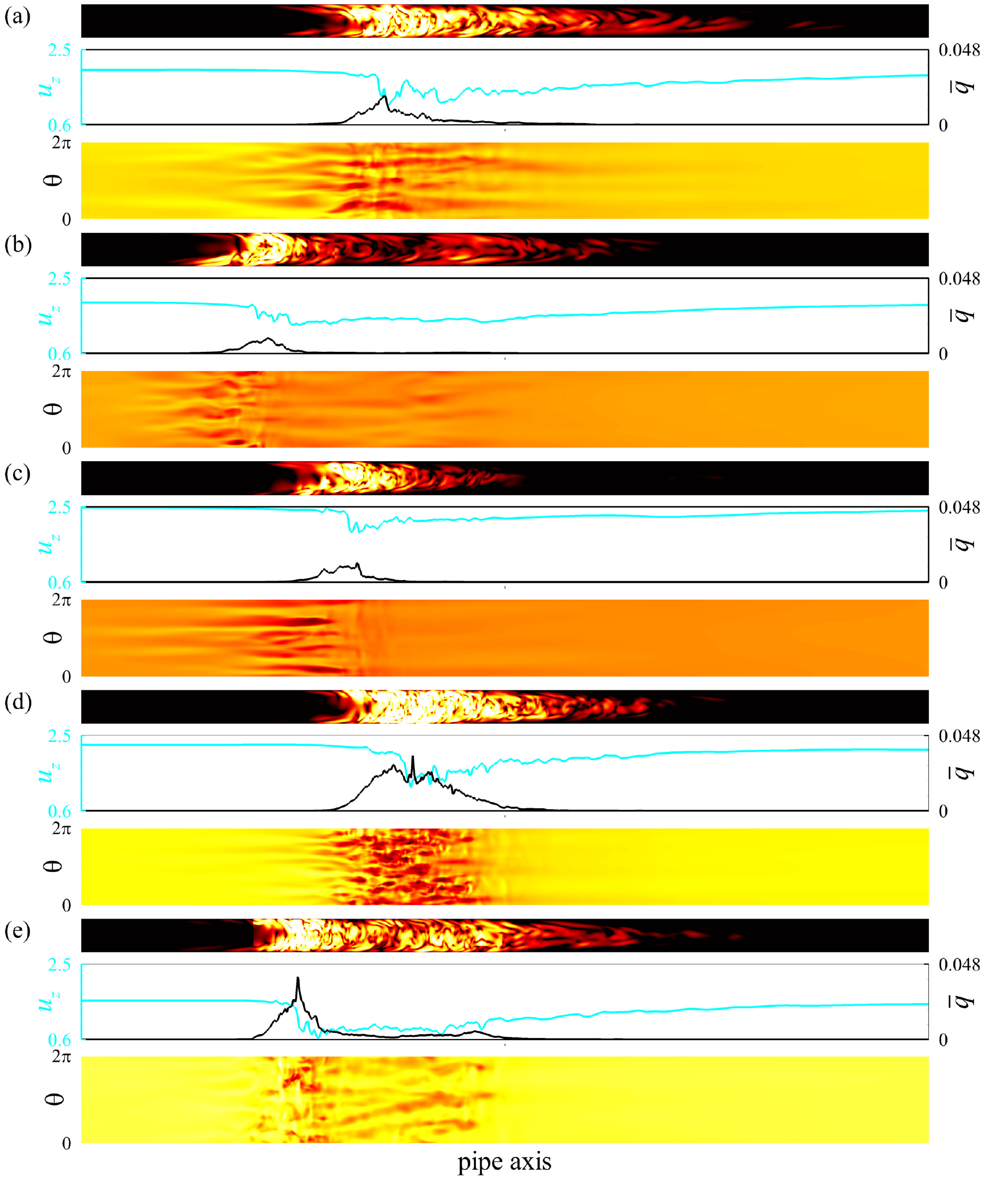}
\caption{\label{fig:puff} (Colour online) Snapshots of a turbulent puff at $\Rey=2040$ in a frame co-moving at the mean speed: (a) steady flow, (b--e) pulsatile flow with  $(\A,\Wo)=(0.4,7.2)$ and $t/\mathcal{T}=$ $0$ (b), $0.25$ (c), $0.5$ (d) and $0.75$ (e). The flow is from left to right, and a portion of the pipe ($40D$ out of a $100D$ pipe) is shown to highlight localized turbulence. For each set of panels, the top panel shows a colour map of the turbulence intensity $q$ in a logarithmic scale, where white (black) corresponds to high (low) turbulent intensity. The middle one shows the corresponding centreline velocity $u_z(z)$ (cyan line) and cross-sectionally averaged turbulence intensity $\bar{q}(z)$ (black line) along the pipe axis. The bottom one shows a colour map of wall shear stress $\tau_{z}$ in a linear scale. For each shown quantity, the same scale is used in all snapshots.}
\end{figure}

A further discrepancy among previous studies concerns the structure of turbulence near the transition threshold. In steady pipe flow at low $\Rey\approx2000$, turbulence is confined to localized patches of constant length \citep[called puffs; see][]{Wygnanski75}. Puffs have a sharp turbulent--laminar upstream interface and a diffuse downstream interface, as shown in the snapshot of figure~\ref{fig:puff}(a), and travel at approximately the mean speed $U_s$. \citet{Trip12} and \citet{Xu17a} reported puffs in their experiments of pulsatile flow, whereas \citet{Stettler86} found that at low $\Wo$ turbulence resembles a puff with flipped upstream and downstream interfaces. Such inverted puffs appeared in their experiments in a phase-locked manner. Phase-locked turbulence was also reported in the experiments of \citet{Iguchi82} at high pulsation amplitudes. 

These discrepancies call for a numerical study of the transition to turbulence in pulsatile flow. Here  we perform direct numerical simulations (DNS) in the parameter regimes investigated experimentally. DNS is a research tool free of natural disturbance, with which perturbations can be implemented in a well-controlled reproducible manner. Furthermore, in experiments of pulsatile pipe flow, measurements are typically taken at a single or a few streamwise locations. By contrast, DNS provide spatio-temporally resolved velocity fields and hence enable a detailed examination of the dynamics of turbulence as it travel downstream along the pipe.

\section{Methods}\label{sec:numerics}

We consider an incompressible viscous fluid driven through a straight pipe of circular cross-section at a pulsatile flow rate.
The instantaneous Reynolds number is
\begin{equation}
\widetilde{\Rey}(t) = \Rey\cdot[1 + \A\cdot \textrm{sin}(2\pi \cdot t/\mathcal{T})].
\label{eq:Re}
\end{equation}
Lengths and velocities are rendered dimensionless with $D$ and  $U_s$, respectively. Consequently, time is made dimensionless with the advective time unit $D/U_s$ and the pulsation period is $\mathcal{T}=\pi \Rey/(2\Wo^2)$. For most simulations, the pulsation amplitude was set to $\A=0.4$, as in the experiments shown in figure~\ref{fig:exp}. The effect of amplitude was then investigated by performing additional simulations at $\A=0.2$. The Navier--Stokes equations were solved with the openpipeflow.org code of \citet{openpipeflow}, which uses primitive variables and a pressure Poisson equation (PPE) formulation with the influence-matrix method to discretize the equations in cylindrical coordinates $(r, \theta, z)$. In a pipe of up to $96\pi D\approx 300D$ in length, simulations were carried out with up to $7680$ ($K=\pm3840$) and 128 ($M=\pm64$) Fourier modes in the periodic axial and azimuthal directions, respectively. In the radial direction, explicit finite differences on nine-point stencils were employed in grids of up to $N=72$ points. The points were clustered densely close to the pipe wall \citep[see][for details]{openpipeflow}.

The dynamics and spatial structure of turbulence were examined by disturbing the laminar pulsatile flow with a pair of streamwise rolls localized within approximately $3D$ \citep[see][for details]{Mellibovsky09}. As a result, a single turbulent puff emerged and was subsequently tracked as it travelled downstream. Figure~\ref{fig:puff}(a) shows a turbulent puff in steady pipe flow at $\Rey=2040$. Its spatial structure can be seen in the colour maps of turbulence intensity $q=u_r^2+u_{\theta}^2$, the axial profiles of their corresponding cross-sectional average $\bar{q}$ and the streamwise velocity $u_z$ at the pipe centre. The signature of the turbulent puff at the pipe wall is characterized by streaky patterns of shear stress $\tau_z$, whose largest fluctuations are mainly concentrated at the upstream interface.

\begin{figure}
	\centering
	\includegraphics[width=0.7\textwidth]{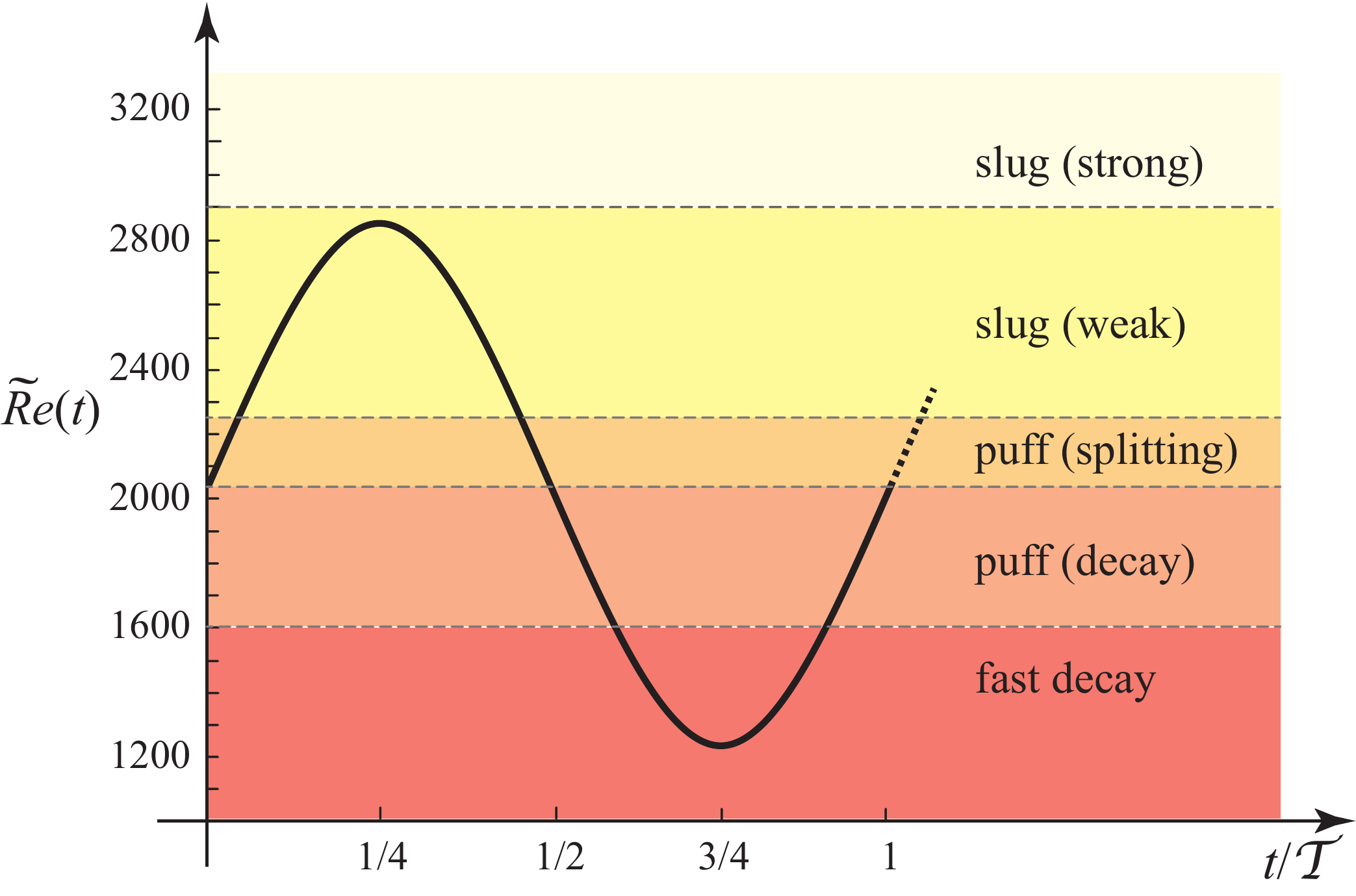}
	\caption{\label{fig:Re} (Colour online) Sketch of the evolution of the instantaneous Reynolds number $\widetilde{\Rey}(t)$ over the pulsation period for $\Rey=2040$ and $A=0.4$ (see \eqref{eq:Re}). Coloured regions depict the different regimes encountered in steady pipe flow as the Reynolds number increases.}
\end{figure}

In order to interpret the ensuing flow dynamics as $\widetilde{\Rey}(t)$ varies through the pulsation cycle, it is useful to briefly summarize the regimes encountered in steady pipe flow (see figure~\ref{fig:Re}). Puffs of constant length occur at low $\Rey\lesssim2250$ and can decay to laminar flow or split, and thereby increase the turbulent fraction. Both processes are stochastic (memoryless) and their competition determines the critical point for the onset of turbulence at $\Rey_c=2040$. For $\Rey<2040$, the decay dynamics are faster and hence dominate, whereas splitting outweighs decay above the critical point \citep{Avila11}. At $\Rey\gtrsim2250$ puffs are superseded by turbulent slugs expanding at constant speed \citep{Barkley15}, and at $\Rey\gtrsim 2900$ slugs develop a sharp downstream interface similar (approximately symmetric) to their upstream interface \citep{Barkley15,Song17}.   

\section{Spatio-temporal dynamics of pulsatile turbulence} \label{sec:localized_turbulence}

\begin{figure}
	\centering
	\includegraphics[width=\textwidth, trim={0cm 0cm 0.5cm 0cm}, clip]{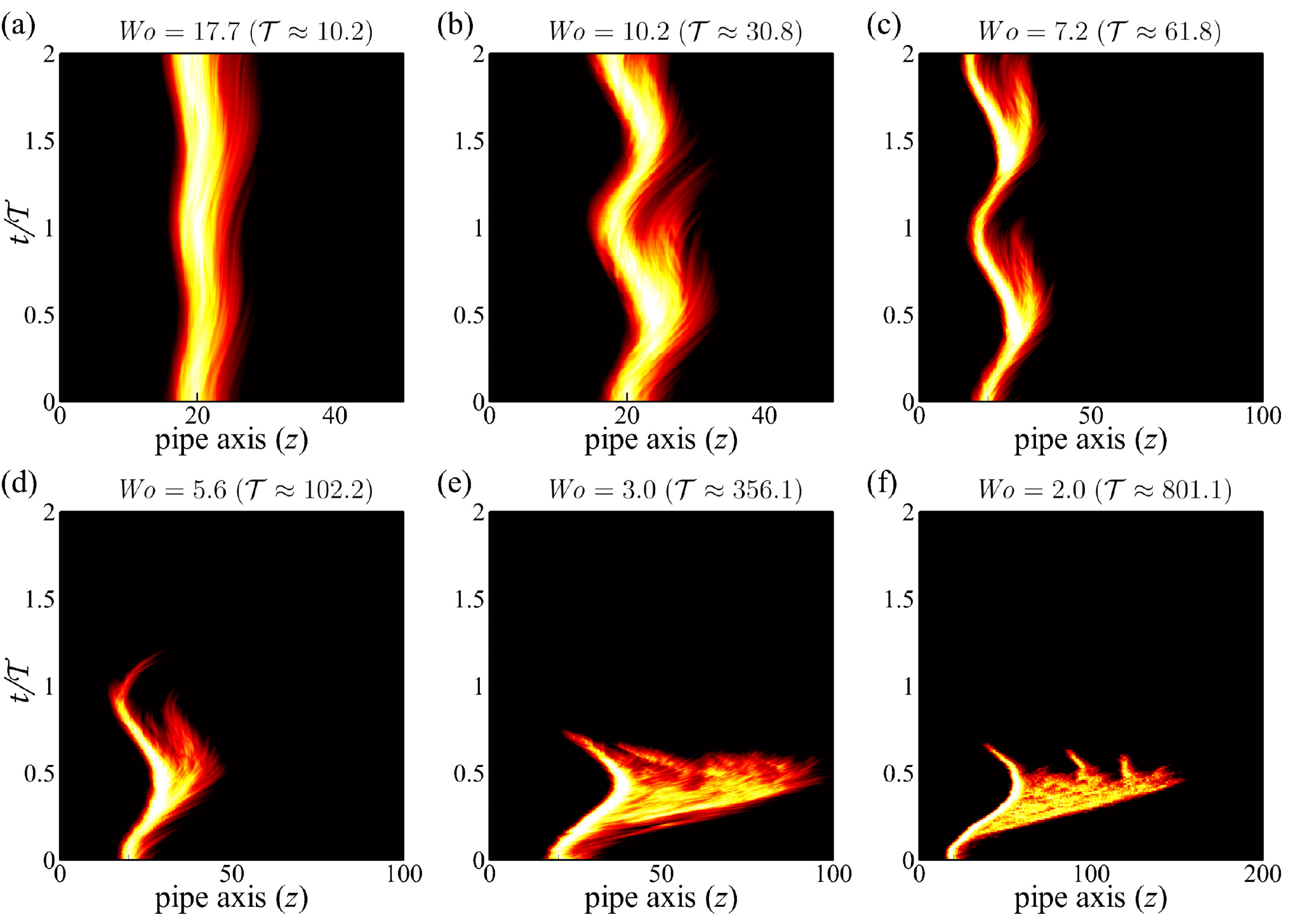}
	\caption{\label{fig:space_time} (Colour online) Space-time diagrams of localized turbulence at $\Rey = 2040$ and $A=0.4$ in a frame co-moving at the mean speed. Colour maps of $\bar{q}(z,t)$ are shown with the same colour scale as in figure \ref{fig:puff}. Note that the pipe length (horizontal axis) was varied with $\Wo$ to avoid interaction of the upstream and downstream interfaces through the imposed periodic boundary conditions.}
\end{figure}

Instantaneous snapshots of a simulation at $\Rey=2040$, $\A=0.4$ and $\Wo=7.2$ are shown in figure~\ref{fig:puff}(b--e) in a frame co-moving with the mean flow speed. Throughout the cycle, the turbulent region grows and shrinks with a certain delay with respect to the instantaneous Reynolds number $\widetilde{\Rey}(t)$, but preserves the characteristic shape of a puff. The spatio-temporal dynamics of the flow is visualized in figure~\ref{fig:space_time}(c) and exhibits incipient stages of puff splitting and decay. Turbulence expands and retreats from the downstream interface, and the propagation speed of the puff oscillates about the mean flow speed. This is not surprising because in steady pipe flow puffs propagate at nearly the mean flow speed at $\Rey=2040$ \citep{Avila11}. As $\Wo$ is reduced, while keeping $\Rey=2040$ and $\A=0.4$ constant, puffs have more time to grow and shrink during the cycle. This leads to splitting and relaminarization, as exemplified in figure~\ref{fig:space_time}(d) for $\Wo=5.6$.  Figure~\ref{fig:space_time}(e--f) shows that by further reducing $\Wo$, the flow responds quasi-statically. Turbulence expands continuously as a slug while $\widetilde{\Rey}(t)$ increases, but collapses irreversibly during the lower half-cycle. By contrast, at high $\Wo$, the turbulent dynamics has little time to react to the rapid pulsation of the flow rate, and the behaviour of steady pipe flow is gradually recovered (see figure~\ref{fig:space_time}a,b). 

\begin{figure}
	\centering
	\includegraphics[width=\textwidth, trim={0cm 0cm 0.5cm 0cm}, clip]{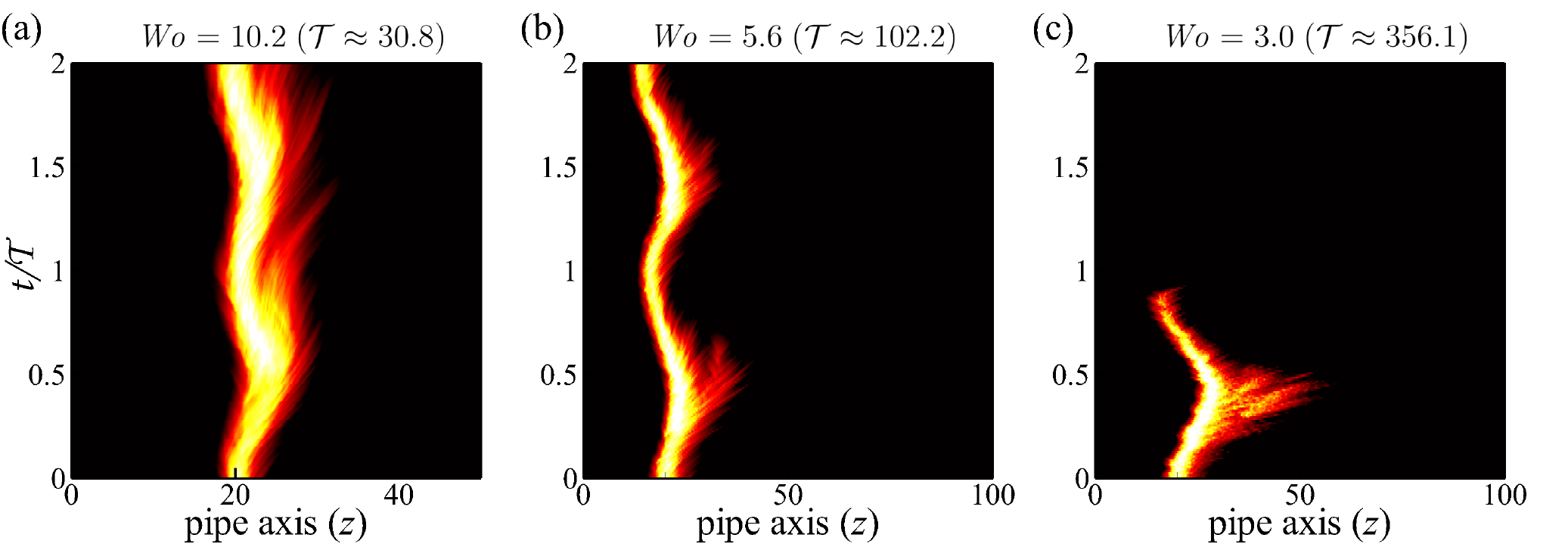}
	\caption{\label{fig:space_time2} (Colour online) Space--time diagrams of localized turbulence at $\Rey = 2040$ and $A=0.2$ in a frame co-moving at the mean speed. The colour maps are as in figure~\ref{fig:space_time}.}
\end{figure}

Figure~\ref{fig:space_time2} shows analogous space-time diagrams also at $\Rey=2040$, but for lower pulsation amplitude $\A=0.2$. Here the instantaneous Reynolds number $\widetilde{\Rey}(t)$ varies across a narrower range and this permits smaller variations in the size of turbulence. This hinders the occurrence of both splitting and relaminarization, enabling turbulence to survive for longer times. Together, the visualizations of figures~\ref{fig:space_time} and \ref{fig:space_time2} suggest that pulsation shifts the transition threshold to larger $\Rey$. Reducing $\Wo$ or increasing $\A$ considerably enhances the stability of the laminar pulsatile flow.

These visualizations also highlight the need to use long pipes in experiments in order to capture the asymptotic dynamics of the system at low pulsation frequency. For instance, let us consider the case $(\Rey,\A,\Wo) = (2040,0.4,2.0)$ shown in figure~\ref{fig:space_time}(f). In the experiments of \citet{Stettler86}, localized turbulence would reach the end of the pipe at $t/\mathcal{T}\approx0.41$ (space was converted into time via the mean flow speed). In their experiments, a long slug would be observed to reach the end of the pipe and $\Rey=2040$ would be taken as turbulent. However, turbulence begins decaying at the downstream interface at $t/\mathcal{T}\approx0.5$ and has collapsed completely by $t/\mathcal{T}\approx0.7$. Hence $\Rey=2040$ is in fact below the transition threshold if a full cycle is considered. 

\begin{figure}
	\centering
	\includegraphics[width=0.85\textwidth, trim={0cm 0.3cm 0cm 0cm}, clip]{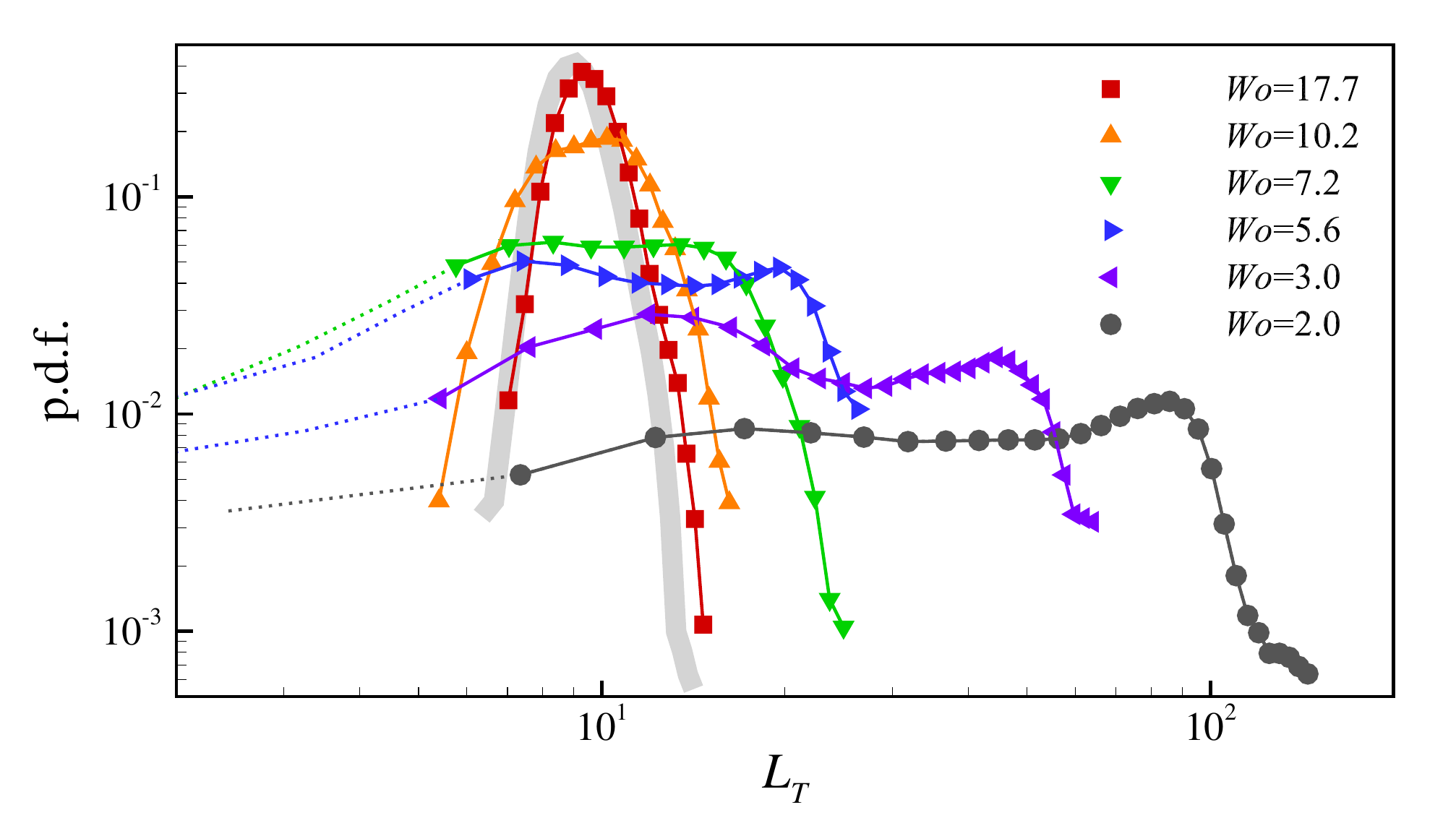}
	\caption{\label{fig:puff_length} (Colour online) Probability density functions (p.d.f.) of turbulent length $L_T$ at $\Rey=2040$, $A=0.4$ and several $\Wo$. The data with $L_T<5$ are shown as dotted lines because turbulence starts to decay irreversibly. Reference data for a puff in steady pipe flow are shown as a grey thick line.}
\end{figure}

The effect of pulsation frequency on the length of the turbulent region $L_T$ was systematically studied by performing 20 runs at each $\Wo$ shown in figure~\ref{fig:space_time}. Here a threshold on the turbulence intensity ($\bar{q}_c=2\times10^{-3}$) was set to distinguish the beginning and end of turbulent flow regions. In the presence of split puffs, $\bar{q}$ may fall below this threshold in the gap between puffs. In such cases, $L_T$ was taken as the length encompassing all puffs. We note that larger (smaller) values of $\bar{q}_c$ result in shorter (longer) turbulent length, but such difference is trivial for the following analysis. Figure \ref{fig:puff_length} shows the probability density function (p.d.f.) of $L_T$. At $\Wo=17.7$, $L_T$ is narrowly distributed about $9.3$ (in steady pipe flow, shown in grey, $L_T\approx9$), and no effect of the pulsation is observed because $\widetilde{\Rey}(t)$ varies too rapidly for the puff to adapt. At $\Wo=10.2$ the length begins to adjust to the flow pulsation and varies between $16$ and $5$, centred around a much broader peak at $10$ in the p.d.f. By further reducing to $\Wo=7.2$, the p.d.f. turns bimodal with a peak at $L_T=8$ corresponding to one puff, and another peak at $14$ corresponding to the incipient stages of splitting (see figure~\ref{fig:space_time}c). At $\Wo=5.6$ splitting events occur often and two clear peaks at $L_T=7$ and $20$, corresponding to one and two puffs, can be clearly discerned. As $\Wo$ is further reduced, the distribution becomes progressively flat because of the continuous expansion of the turbulent slug (see e.g. figure~\ref{fig:space_time}e). Note that when $L_T\lesssim5$ puffs begin to decay irreversibly and the flow fully relaminarizes.

\begin{figure}
	\centering
	\includegraphics[width=\textwidth, trim={0.8cm 0cm 2.1cm 0cm}, clip]{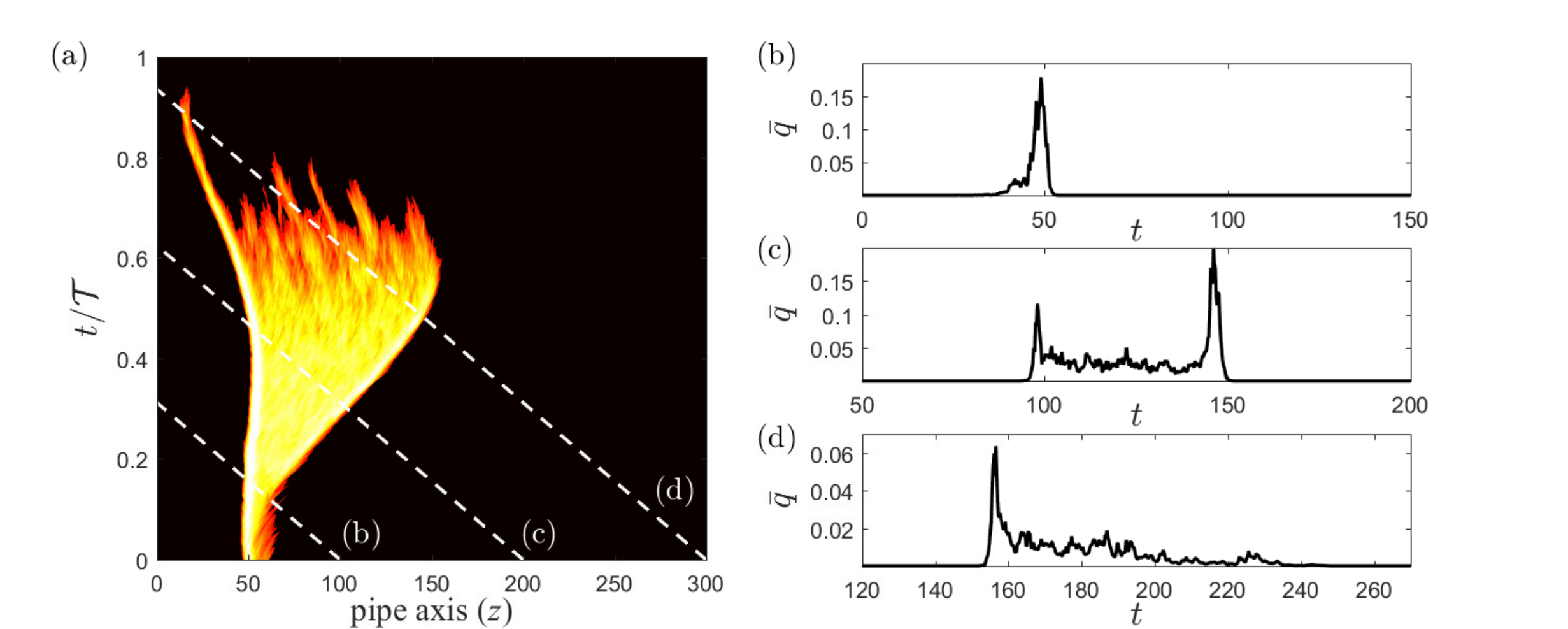}
	\caption{\label{fig:inverted_puff} (Colour online) (a) Space--time diagram of localized turbulence at $(\Rey,\A,\Wo)=(2500,0.4,3.5)$ in a $300D$ pipe in a frame co-moving at the mean flow speed. The dashed lines indicate fixed downstream locations in the pipe. (b--d) Time series of the cross-sectionally averaged turbulence intensity $\bar{q}$ at the downstream locations marked in (a).}
\end{figure}

Figure~\ref{fig:inverted_puff}(a) shows a space--time diagram obtained at $\Rey=2500$, $\A=0.4$ and $\Wo=3.5$, where the dashed lines denote the position of three (virtual) measurement points at fixed downstream locations. The measured time series of $\overline{q}$ are shown in figure~\ref{fig:inverted_puff}(b--d) and are analogous to time traces measured in experiments.  Initially, a puff forms (figure~\ref{fig:inverted_puff}b) and quickly begins to grow as a slug. As $\widetilde{\Rey}(t)>2900$ the downstream interface becomes increasingly sharp and the slug appears nearly symmetric (figure~\ref{fig:inverted_puff}c). As $\widetilde{\Rey}(t)$ decreases, the structure begins to shrink and finally collapses entirely. A stationary observer at $z=300$ (figure~\ref{fig:inverted_puff}d) sees first the passage of the sharp downstream interface, while at the time that the rear part of the structure reaches the measurement point, turbulence is monotonically decaying. As a result,  a diffuse upstream interface is seen. The signal thus resembles an inverted puff, thereby explaining the observation of \cite{Stettler86}, but corresponds in fact to the decay of a slug in pulsatile pipe flow. Note also that if a continuous disturbance as the orifice of \cite{Stettler86} were used, then such turbulent slugs would be observed periodically at the measurement point (phase-locked turbulence).

\section{Lifetimes of localized turbulence in pulsatile pipe flow}

\begin{figure}
\centering
\includegraphics[width=\textwidth, trim={0cm 0.4cm 0cm 0cm},clip]{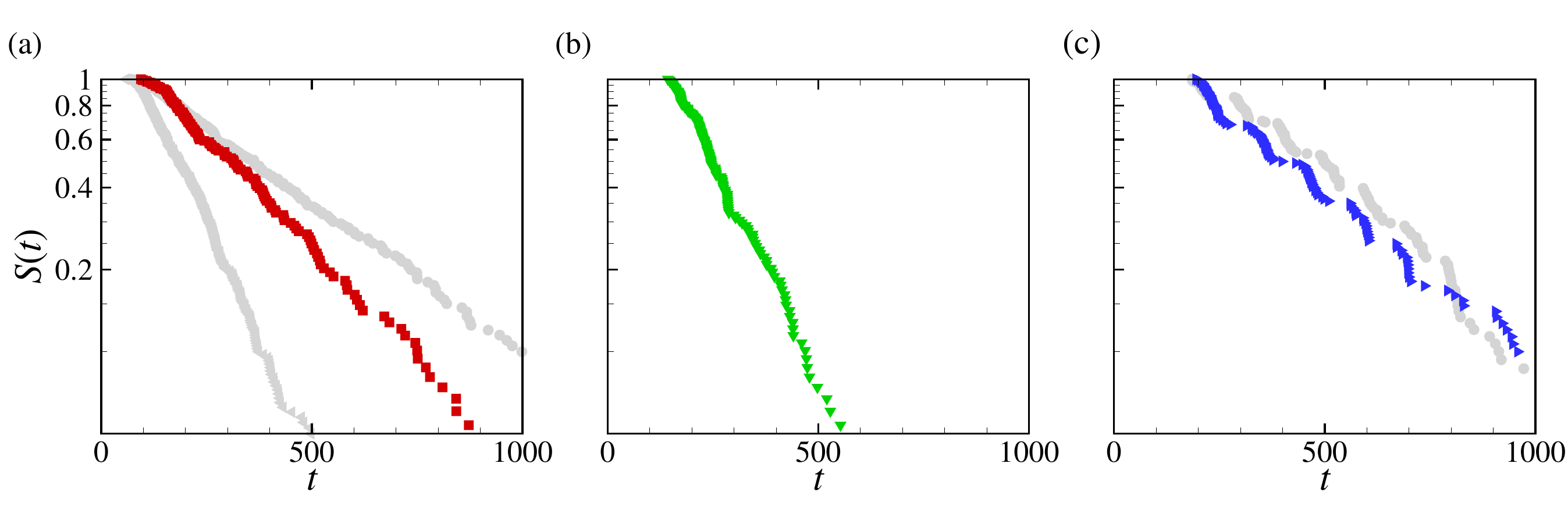}
\caption{\label{fig:St} (Colour online) Survival probability $S(t)$ of localized turbulence at $(\Rey,\A, \Wo)=(1820,0.4,17.7)$, $(2040,0.4,7.2)$ and $(2320,0.4,5.6)$, respectively. In (a), grey triangles (circles) show the survival probability at $\Rey=1820$ ($\Rey=1860$) in steady pipe flow \citep{Avila10}. In (c), grey circles show the survival probability at $(\Rey,\A,\Wo)=(2040,0.2,5.6)$.}
\end{figure}

\begin{table}
	\centering
	\begin{tabular}{lccccccccccc}
		$\Rey$ & $A$ & $\Wo$ & $\mathcal{T}$ & $\tau^*$ & $\tau^*/\mathcal{T}$ & $t^*_0$ & $t^*_0/\mathcal{T}$ & $P(L=350)$ & $P_\text{exp}(L=350)$\\ \addlinespace[2mm]
		$1820$ & $0.4$ & $17.7$ & $9.1$ & $321.5$ & $35.2$ & $102.0$ & $11.2$ & $0.48\pm0.02$ & $0.46\pm0.02$\\
		$2040$ & $0.4$ & $7.2$ & $61.8$ &  $244.0$ & $3.9$ & $142.9$ & $2.3$ & $0.19\pm0.02$ & $0.20\pm0.02$\\
		$2320$ & $0.4$ & $5.6$ & $116.2$ & $353.6$ & $3.0$ & $186.5$ & $1.6$ & $0.51\pm0.02$ & $0.49\pm0.02$ \\
		$2040$ & $0.2$ & $5.6$ & $102.2$ & $429.5$ & $4.2$ & $185.4$ & $1.8$ & $0.55\pm0.02$ & --- \\
	\end{tabular}
	\caption{\label{tab:DNS} Lifetime statistics of localized turbulence in pulsatile pipe flow. Here $\tau^*$ is the median lifetime and $t^*_0$ is time of the first relaminarization. The survival probability at $L=350$ from our simulations is compared to the survival probability measured experimentally at $L=350$ by \citet{Xu17a}. }
\end{table}

\citet{Xu17a} introduced two novelties with respect to previous studies of transition in pulsatile pipe flow. First, they disturbed the flow with an injection, which allowed triggering turbulence at lower $\Rey$ than the orifice of \citet{Stettler86,Trip12}. Their injection generated a single turbulent puff and is similar  to the localized pair of rolls used here to disturb the flow. Second, at a fixed $\Wo$, they measured the survival probabilities of turbulent puffs $P_\text{exp}(L)$, consisting of the number of puffs detected at distance $L$ downstream of the injection, divided by the total number of realizations. This was inspired by previous studies of steady Couette and pipe flows, where the underlying relaminarization process is memoryless \citep{bottin1998a,Faisst04}. More specifically, the probability of a puff surviving beyond time $t$ in steady pipe flow is given by the survivor function $S(t) = \textrm{exp}[-(t-t_0)/\tau]$, where $\tau$ is the mean lifetime and $t_0$ is the initial formation time of the puff after the laminar flow is disturbed \citep{Hof06}. The key point here is that the dependence of initial condition (disturbance method) is fully contained in $t_0$, whereas $\tau$ is uniquely determined by $\Rey$~\citep[see][]{deLozar09}. Hence stability thresholds determined with survival probabilities are intrinsic to the flow (disturbance independent so long as a single turbulent puff is generated). \citet{Xu17a} defined the transition threshold $\Rey_T(\Wo)$ such that $P_\text{exp}(L)=0.5$. They performed measurement with $L=350$, $1300$ and $2250$, as indicated by the three different triangle styles shown in figure~\ref{fig:exp} (in their experiments, $L$ was increased, as $\Wo$ was decreased). 

We investigated the effect of pulsation on turbulence relaminarization at three selected points in $(\Rey,\Wo)$ parameter space to allow a direct comparison to \citet{Xu17a} for $\A=0.4$. In each case 150 runs were performed in a pipe of length $100$ to sufficiently resolve the survivor functions. At high frequency, $\Wo=17.7$ and $\Rey=1820$, the relaminarization process is memoryless as shown by the corresponding survivor function (red squares) in figure~\ref{fig:St}(a). In comparison to steady pipe flow (grey triangles), the survival probability slightly increases but remains below that of steady flow at $\Rey=1860$ (grey circles). Regarding lifetimes, the effect of pulsation at $\Wo=17.7$ is equivalent to a small shift (less than $2\%$) in Reynolds number.

At $\Wo=7.2$ the relaminarization process remains approximately memoryless (figure~\ref{fig:St}b), but this character is lost as the pulsation frequency is further reduced to $\Wo=5.6$ (figure~\ref{fig:St}c). The survivor function is characterized by constant steps over a half-cycle (during which no relaminarization occurs), followed by exponential decay during the subsequent half-cycle. Table~\ref{tab:DNS} summarizes the results of our simulations and includes a direct comparison to the experiments of \citet{Xu17a}. Despite the uncertainty in $t^*_0$ and the  limited sample sizes, there is excellent quantitative agreement between the two datasets. The effect of pulsation amplitude was tested by computing lifetime statistics at $(\Rey,\A,\Wo)=(2040, 0.2, 5.6)$. The corresponding survivor function is shown in grey circles in figure~\ref{fig:St}(c) and is very similar to that of $(\Rey,\A,\Wo)=(2320, 0.4, 5.6)$, shown as blue triangles. Thus, doubling the pulsation amplitude from $\A=0.2$ to $0.4$ requires an increase of the Reynolds number by $14\%$ to preserve similar relaminarization dynamics and probability. This clearly demonstrates the stabilizing effect of increasing pulsation amplitude. 

\section{Assessment of experimentally measured transition thresholds}

\begin{figure}
	\centering
	\includegraphics[width=0.8\textwidth, trim={0.3cm 0.5cm 0.2cm 0cm},clip]{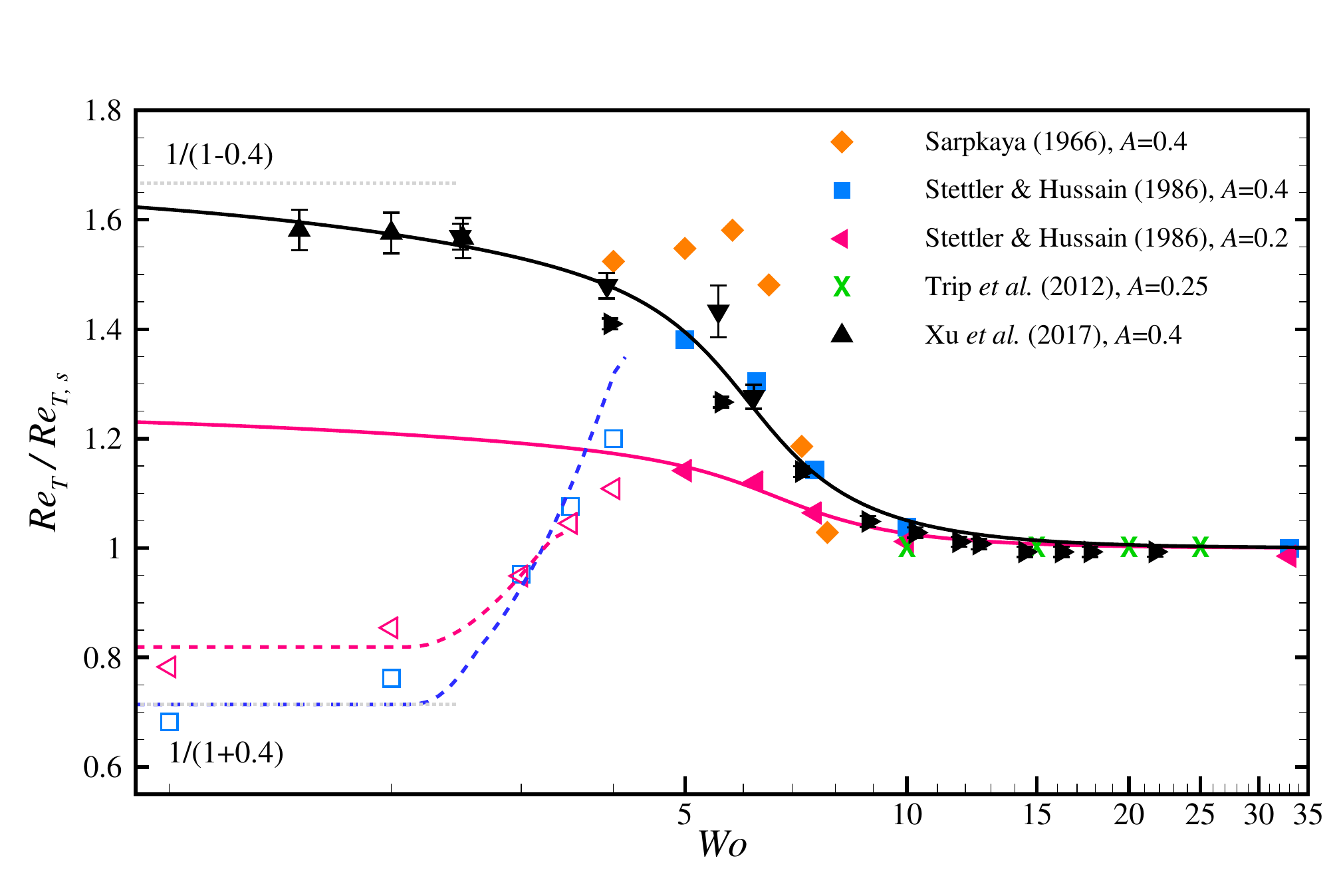}
	\caption{\label{fig:model} (Colour online) Experimentally measured transition thresholds from experiments rescaled by the respective steady thresholds. The solid and dashed lines show the threshold given by  \eqref{eq:long_pipe_model} and \eqref{eq:short_pipe_time}, respectively. In a sufficiently long pipe,  $\Rey_T/\Rey_{T,s}$ approaches the upper limit $1/(1-A)$ as $\Wo$ approaches zero, whereas for insufficient length $1/(1+A)$ is approached. Both limits are shown as thin grey dotted lines for $\A=0.4$. }
\end{figure}

\citet{Xu17a} proposed that, in pulsatile pipe flow at low amplitude $\A\le 0.4$, transition can be divided into three regimes according to the pulsation frequency. These regimes are summarized and analysed here in view of the new insights gained from our numerical simulations. 

\subsection{High pulsation frequencies}

There is consensus between the experiments of \citet{Stettler86, Trip12} and \citet{Xu17a} and our simulations that, as the pulsation frequency increases, the dynamics of steady pipe flow are recovered. When their experimentally measured thresholds shown in figure~\ref{tab:exp} are normalized with the respective values of the steady case, their datasets collapse together (see figure~\ref{fig:model}). This confirms that the difference in their measured transition thresholds is simply due to the different disturbance methods and turbulence detection criteria employed. The data of \citet{Xu17a} were obtained with three different measurement lengths $L$, and each dataset was normalized here by its corresponding $\Rey_{T,s}$ (see table~\ref{tab:exp}). This is necessary because their criterion to determine the threshold, namely $P_\text{exp}(L)=0.5$, depends on $L$. Longer pipes result in smaller survival probability, if all parameters are kept constant. The data of \citet{Sarpkaya66}  approach the transition threshold of the steady case as $\Wo$ increases, but his measurements reach only up to $\Wo=8$.

\citet{Peacock98} reported instead that the flow is monotonically stabilized as the frequency of the pulsation increases. Unfortunately, in the way that the data are presented in their paper, it is not possible to extract the exact parameter values at which experiments were performed. However, their measurements were mostly performed with $\A>1$, which suggests that transition at large amplitude may be significantly different. This hypothesis is supported by the experiments of \citet{Lodahl98}, who also focused on large amplitudes.

\subsection{Intermediate pulsation frequencies}

The experiments of \citet{Sarpkaya66,Stettler86} and \citet{Xu17a} and our simulations also agree that, as $\Wo$ is decreased, transition to turbulence is shifted to higher Reynolds numbers. In addition, once the dependence on the disturbance and detection criteria is left out, the data of \citet{Stettler86} and \citet{Xu17a} are in excellent quantitative agreement down to $\Wo=5$. The data of \citet{Sarpkaya66} exhibit a much higher degree of stabilization. He perturbed the flow at a fixed phase and measured the lengths of turbulent regions $L_T$ at three downstream locations separated by $370D$. He marked the flow as turbulent if $L_T$ was approximately constant at all the three measurement points. However, the space-time diagrams of figure~\ref{fig:space_time} show that $L_T$ varies with the phase and exhibits large fluctuations for low Womersley numbers $\Wo \lesssim 7$ (see figure~\ref{fig:puff_length}). Hence we believe that his criterion resulted in an inaccurate determination of the transition threshold.

\subsection{Low pulsation frequencies: experiments in long pipes}

In the experiments of \citet{Xu17a}, the transition threshold continues to shift to larger Reynolds number as $\Wo$ is reduced. This effect gradually saturates, and they explained this by considering the quasi-steady limit ($\Wo\rightarrow 0$) in a sufficiently long pipe. They argued that the flow should remain turbulent provided that the instantaneous Reynolds number remained above the transition threshold for steady flow throughout the cycle, i.e.\ $\min\widetilde{\Rey}(t)\ge\Rey_{T,s}$. This requirement yields the threshold 
\begin{equation}
\Rey_{T}(\Wo=0)=\dfrac{\Rey_{T,s}}{1-A}.
\end{equation}
\citet{Xu17a} showed not only that their data approaches this limit, but also that, for small $\Wo\lesssim 3$, survival probabilities can be approximated from the corresponding steady-state values. The thresholds measured by \citet{Xu17a} are well approximated by the empirical correlation
\begin{equation}
\Rey_T(\Wo)= \Rey_{T,s}\left[1+\dfrac{A}{2(1-A)}\left(1+\dfrac{2}{\pi}\arctan\left({\alpha^2\Wo^{-1}-\beta^2\Wo^3}\right)\right)\right],
\label{eq:long_pipe_model}
\end{equation}
which guarantees by construction that the correct transition thresholds are recovered in both the low- and high-frequency limits, independently of the value of the fitting parameters (here $\alpha\approx2.08$ and $\beta\approx0.07$ obtained with a nonlinear least-squares fit to their data). The same value of the fitting parameters renders good approximations of the data of \citet{Stettler86} for $A=0.4$ and $0.2$.

\subsection{Low pulsation frequencies: prediction of the transition threshold in short pipes}

The quasi-steady limit proposed by \citet{Xu17a} relies on the assumption of a sufficiently long pipe subject to a single impulsive disturbance. However, the pulsation period diverges as $1/\Wo^2$, which ultimately prevents the observation of turbulence over a full cycle within the available measurement length. Even in their very long pipe with $L=2250$, the minimum feasible Womersley number is $\Wo^*\approx 1.5$, whereas the pipe of \citet{Stettler86} with $L=330$ has $\Wo^*\approx 4$. In many practical situations, lengths are even shorter and disturbances are present all the time. Here we propose a prediction of the transition threshold in continuously disturbed finite-length pipes. 

\citet{Stettler86} observed that, at low $\Wo\lesssim 5$, turbulence appeared in a phase-locked manner, when the instantaneous Reynolds number exceeded the transition Reynolds number of the steady case, i.e.\ $\max\widetilde{\Rey}(t)\ge\Rey_{T,s}$. From this they argued that, in the quasi-steady limit, the transition threshold should be given by
\begin{equation}
\Rey_{T}(\Wo=0)=\dfrac{\Rey_{T,s}}{1+A},
\end{equation}
and this prediction was found to be in good agreement with their experiments. According to this argument, turbulence is first triggered at time $t_\text{tr}$, with $\widetilde{\Rey}(t_\text{tr})=\Rey_{T,s}$. By plugging this in equation~\eqref{eq:Re}, we obtain
\begin{equation}
 t_\text{tr}= \dfrac{\mathcal{T}}{2\pi} \sin^{-1}\left[\dfrac{1}{A}\left(\dfrac{\Rey_{T,s}}{\Rey}-1\right)\right].
 \label{eq:t_T}
\end{equation}
Subsequently, turbulence travels downstream until it either decays or is flushed at the end of the pipe. For simplicity, we assume that turbulence decays when $\widetilde{\Rey}(t_\text{d})=\Rey_{D,s}\approx 1600$, at which point the lifetime of a puff in steady pipe flow becomes negligible \citep{Hof08}. This yields the following decay time:
\begin{equation}
 t_\text{d}= \dfrac{\mathcal{T}}{2\pi} \sin^{-1}\left[\dfrac{1}{A}\left(\dfrac{\Rey_{D,s}}{\Rey}-1\right)\right].
\label{eq:t_D}
\end{equation}
For $(\Rey,\A,\Wo) = (2040,0.4,2)$, we obtain $t_d\approx 473$ ($t_d/\mathcal{T}\approx 0.6$), which is in good agreement with our simulation of figure~\ref{fig:space_time}(f).

At a measurement point downstream of the disturbance, turbulence will be detected if the time lapsed between the generation and decay of turbulence is longer than its travel time. Otherwise, turbulence will decay between disturbance and measurement point and laminar flows will be detected. Hence the measured transition threshold should be given by the nonlinear equation  
\begin{equation}
\dfrac{L}{\overline{c}}= t_\text{d} - t_\text{tr},
\label{eq:short_pipe_time}
\end{equation}
where $L$ is the distance between disturbance and measurement point and $\overline{c}$ is the average propagation speed of turbulence during the time interval $t\in [t_\text{tr}, t_\text{d}]$. By assuming that the instantaneous propagation speed of turbulence is equal to the instantaneous bulk speed, direct integration yields
\begin{equation}
\overline{c} = \dfrac{1}{t_\text{d}- t_\text{tr}} \int_{t_\text{tr}}^{t_\text{d}}\left[1 + A\sin\left(\dfrac{2\pi t}{\mathcal{T}}\right)\right] \text{d}t = 1 + \dfrac{\mathcal{T}A}{2\pi(t_\text{d}- t_\text{tr})}\left[\cos\left(\dfrac{2\pi t_\text{tr}}{\mathcal{T}}\right)-\cos\left(\dfrac{2\pi t_\text{d}}{\mathcal{T}}\right) \right].
\end{equation}
The dashed lines in figure~\ref{fig:model} show the solution of \eqref{eq:short_pipe_time} for $\A=0.2$ and $\A=0.4$. Both curves capture qualitatively the trend of the transition threshold measured by \citet{Stettler86}.  The agreement could be made more quantitative by computing $\overline{c}$ from the actual $\Rey$-dependent speed of downstream laminar--turbulent interfaces in steady pipe flow \citep{Barkley15}. Finally, we note that the choice $\Rey_{D,s}\approx 1600$ does not significantly affect the prediction. In fact, replacing $1600$ by $1700$ or $1800$ cannot be noticed in the scale used in figure~\ref{fig:model}.

\section{Conclusions}{\label{sec:discussion}}

Pulsatile pipe flow is linearly unstable, but in experiments transition to turbulence is observed well below the linear stability thresholds reported by \citet{Thomas11}. Hence finite-amplitude disturbances are required to trigger turbulence and transition thresholds are disturbance-dependent. Once this dependence is removed (by normalizing thresholds with respect to steady pipe flow), there is very good agreement in the thresholds measured by \citet{Stettler86}, \citet{Trip12} and \citet{Xu17a} at high frequencies. These authors and \citet{Sarpkaya66} already pointed out that, at sufficiently high $\Wo$, the dynamics of steady pipe flow is rapidly recovered. At $\Wo=17.7$, only a slight modulation can be detected in the propagation speed of turbulent puffs and their lifetimes are consistent with a shift in $\Rey$ of less than $2\%$ with respect to the steady case. 

As $\Wo$ is reduced, transition is substantially delayed to higher $\Rey$ and the experiments of \cite{Stettler86} and \citet{Xu17a} agree quantitatively down to intermediate frequencies $\Wo\approx 5$. In this regime, our lifetimes statistics are found to be in excellent quantitative agreement with those measured by \cite{Xu17a}. Because of computational constraints, it is not feasible to compute lifetimes at lower frequencies. \cite{Stettler86} reported that, for $\Wo\lesssim 5$, the transition threshold dropped dramatically, whereas \cite{Xu17a} reported a progressive stabilization of the laminar flow down to $\Wo\approx1.5$. \cite{Xu17a} suggested that, in a sufficiently long pipe, the transition threshold is $\Rey_{T,s}/(1-\A)$ in the limit $\Wo\rightarrow 0$. Their data were in very good agreement with a quasi-steady approximation of the lifetime statistics. Furthermore, they argued that the pipe of \cite{Stettler86} was too short to permit following the evolution of turbulence over a full pulsation cycle for $\Wo<\Wo^*=4$. In this paper, we obtained a prediction of the transition threshold for this case, and showed that it qualitatively captures the thresholds measured by \cite{Stettler86} at low frequencies. In the limit $\Wo\rightarrow 0$, the predicted transition threshold is $\Rey_{T,s}/(1+\A)$ and this is in excellent agreement with their measurements at $Wo=1$. We also  showed that if $\Wo<\Wo^*$ in experiments, then phase-locked turbulence is observed, and this explains also why \cite{Stettler86} reported `inverted puffs', which were in fact decaying slugs.

From the experiments of \cite{Stettler86} and \citet{Xu17a},  and our simulations, a comprehensive picture of the transition for low pulsation amplitudes $\A\le0.4$ emerges. Here the flow does not substantially differ from laminar parabolic Poiseuille flow, and so the transition scenario can be fully interpreted in terms of the steady case. For large pulsation amplitude, \citet{Peacock98} and \cite{Lodahl98} reported a monotonic stabilization of the flow as $\Wo$ increases. At large amplitude, the oscillatory flow component is much larger than the steady component and the flow is far from parabolic. The flow profile develops inflection profiles, which lead to a linear instability at large $\Rey$ \citep{Thomas11}. This instability is present in the purely oscillatory case and is also subcritical \citep{Feldmann16}. It emerges from the Stokes layer and is thus governed by other physical mechanisms. The study of transition at intermediate amplitudes, where both mechanisms are expected to compete, remains a challenge of physiological relevance to be addressed in the future. 

\section*{Acknowledgment}

We thank Dr.~Baofang Song, Dr.~Kerstin Avila and Dr.~Daniel Feldmann for fruitful discussions. D.X.\ gratefully acknowledges the support from the Alexander von Humboldt Foundation (3.5-CHN/1154663STP).  We thank the Regionales Rechenzentrum Erlangen of the Friedrich-Alexander-Universit{\"a}t Erlangen-N{\"u}rnberg for providing computing time.

\bibliographystyle{jfm}

\bibliography{PPF}

\begin{thebibliography}{25}
\expandafter\ifx\csname natexlab\endcsname\relax\def\natexlab#1{#1}\fi

\bibitem[Avila {\em et~al.\/}(2011)Avila, Moxey, de~Lozar, Avila, Barkley \&
  Hof]{Avila11}
{\sc Avila, K., Moxey, D., de~Lozar, A., Avila, M., Barkley, D. \& Hof, B.}
  2011 The onset of turbulence in pipe flow. {\em Science\/} {\bf 333},
  192--196.

\bibitem[Avila {\em et~al.\/}(2010)Avila, Willis \& Hof]{Avila10}
{\sc Avila, M., Willis, A.~P. \& Hof, B.} 2010 On the transient nature of
  localized pipe flow turbulence. {\em J. Fluid Mech.\/} {\bf 646}, 127--136.

\bibitem[Barkley(2016)]{Barkley16}
{\sc Barkley, D.} 2016 Theoretical perspective on the route to turbulence in a
  pipe. {\em J. Fluid Mech.\/} {\bf 803}, P1.

\bibitem[Barkley {\em et~al.\/}(2015)Barkley, Song, Mukund, Lemoult, Avila \&
  Hof]{Barkley15}
{\sc Barkley, D., Song, B., Mukund, V., Lemoult, G., Avila, M. \& Hof, B.} 2015
  The rise of fully turbulent flow. {\em Nature\/} {\bf 526}, 550--553.

\bibitem[Bottin \& Chat\'e(1998)]{bottin1998a}
{\sc Bottin, S. \& Chat\'e, H.} 1998 {Statistical analysis of the transition to
  turbulence in plane Couette flow}. {\em Eur. Phys. J. B\/} {\bf 6}~(1),
  143--155.

\bibitem[Chiu \& Chien(2011)]{chiu2011}
{\sc Chiu, J.-J. \& Chien, S.} 2011 Effects of disturbed flow on vascular
  endothelium: pathophysiological basis and clinical perspectives. {\em
  Physiol. Rev.\/} {\bf 91}~(1), 327--387.

\bibitem[Faisst \& Eckhardt(2004)]{Faisst04}
{\sc Faisst, H. \& Eckhardt, B.} 2004 Sensitive dependence on initial
  conditions in transition to turbulence in pipe flow. {\em J. Fluid Mech.\/}
  {\bf 504}, 343--352.

\bibitem[Feldmann \& Wagner(2016)]{Feldmann16}
{\sc Feldmann, D. \& Wagner, C.} 2016 {On the influence of computational domain
  length on turbulence in oscillatory pipe flow}. {\em Int. J. Heat Fluid
  Fl.\/} {\bf 61}, 229--244.

\bibitem[Freis \& Heath(1964)]{Freis64}
{\sc Freis, E.~D. \& Heath, W.~C.} 1964 Hydrodynamics of aortic blood flow.
  {\em Circ. Res.\/} {\bf 14}, 105--116.

\bibitem[Hof {\em et~al.\/}(2008)Hof, de~Lozar, Kuik \& Westerweel]{Hof08}
{\sc Hof, B., de~Lozar, A., Kuik, D.~J. \& Westerweel, J.} 2008 Repeller or
  attractor? selecting the dynamical model for the onset of turbulence in pipe
  flow. {\em Phys. Rev. Lett.\/} {\bf 101}, 214501.

\bibitem[Hof {\em et~al.\/}(2006)Hof, Westerweel, Schneider \& Eckhardt]{Hof06}
{\sc Hof, B., Westerweel, J., Schneider, T.~M. \& Eckhardt, B.} 2006 Finite
  lifetime of turbulence in shear flows. {\em Nature\/} {\bf 443}, 59.

\bibitem[Iguchi \& Ohmi(1982)]{Iguchi82}
{\sc Iguchi, M. \& Ohmi, M.} 1982 {Transition to turbulence in a pulsatile pipe
  flow: Part 2, Characteristics of reversing flow accompanied by
  relaminarization}. {\em Bulletin of JSME\/} {\bf 25}, 1529--1536.

\bibitem[Lodahl {\em et~al.\/}(1998)Lodahl, Sumer \& Freds{\o}e]{Lodahl98}
{\sc Lodahl, C.~R., Sumer, B.~M. \& Freds{\o}e, J.} 1998 Turbulent combined
  oscillatory flow and current in a pipe. {\em J. Fluid Mech.\/} {\bf 373},
  313--348.

\bibitem[de~Lozar \& Hof(2009)]{deLozar09}
{\sc de~Lozar, A. \& Hof, B.} 2009 An experimental study of the decay of
  turbulent puffs in pipe flow. {\em Phil. Trans. R. Soc. A\/} {\bf 367},
  589--599.

\bibitem[Mellibovsky {\em et~al.\/}(2009)Mellibovsky, Meseguer, Schneider \&
  Eckhardt]{Mellibovsky09}
{\sc Mellibovsky, F., Meseguer, A., Schneider, T.~M. \& Eckhardt, B.} 2009
  Transition in localized pipe flow turbulence. {\em Phys. Rev. Lett.\/} {\bf
  103}, 0540502.

\bibitem[Peacock {\em et~al.\/}(1998)Peacock, Jones, Tock \& Lutz]{Peacock98}
{\sc Peacock, J., Jones, T., Tock, C. \& Lutz, R.} 1998 The onset of turbulence
  in physiological pulsatile flow in a straight tube. {\em Exp. Fluids\/} {\bf
  24}, 1--9.

\bibitem[Peixinho \& Mullin(2007)]{Peixinho07}
{\sc Peixinho, J. \& Mullin, T.} 2007 Finite-amplitude thresholds for
  transition in pipe flow. {\em J. Fluid Mech.\/} {\bf 582}, 169--178.

\bibitem[Sarpkaya(1966)]{Sarpkaya66}
{\sc Sarpkaya, T.} 1966 {Experimental determination of the critical Reynolds
  number for pulsating Poiseuille flow}. {\em J. Fluids Eng.\/} {\bf 88},
  589--598.

\bibitem[Song {\em et~al.\/}(2017)Song, Barkley, Hof \& Avila]{Song17}
{\sc Song, B., Barkley, D., Hof, B. \& Avila, M.} 2017 Speed and structure of
  turbulent fronts in pipe flow. {\em J. Fluid Mech.\/} {\bf 813}, 1045--1059.

\bibitem[Stettler \& Hussain(1986)]{Stettler86}
{\sc Stettler, J.~C. \& Hussain, A. K.~M.} 1986 On transition of the pulsatile
  pipe flow. {\em J. Fluid Mech.\/} {\bf 170}, 169--197.

\bibitem[Thomas {\em et~al.\/}(2011)Thomas, Bassom, Blennerhassett \&
  Davies]{Thomas11}
{\sc Thomas, C., Bassom, A.~P., Blennerhassett, P.~J. \& Davies, C.} 2011 {The
  linear stability of oscillatory Poiseuille flow in channels and pipes}. {\em
  Phil. Trans. R. Soc. A\/} {\bf 467}, 2643--2662.

\bibitem[Trip {\em et~al.\/}(2012)Trip, Kuik, Westerweel \& Poelma]{Trip12}
{\sc Trip, R., Kuik, D.~J., Westerweel, J. \& Poelma, C.} 2012 An experimental
  study of transitional pulsatile pipe flow. {\em Phys. Fluids\/} {\bf 24},
  014103.

\bibitem[Willis(2017)]{openpipeflow}
{\sc Willis, A.~P.} 2017 The {O}penpipeflow {N}avier--{S}tokes solver. {\em
  SoftwareX\/} {\bf 6}, 124--127.

\bibitem[Wygnanski {\em et~al.\/}(1975)Wygnanski, Sokolov \&
  Friedman]{Wygnanski75}
{\sc Wygnanski, I.~J., Sokolov, M. \& Friedman, D.} 1975 {On transition in a
  pipe. Part 2. The equilibrium puff}. {\em J. Fluid Mech.\/} {\bf 69},
  283--304.

\bibitem[Xu {\em et~al.\/}(2017)Xu, Warnecke, Song, Ma \& Hof]{Xu17a}
{\sc Xu, D., Warnecke, S., Song, B., Ma, X. \& Hof, B.} 2017 Transition to
  turbulence in pulsating pipe flow. {\em J. Fluid Mech.\/} {\bf 831},
  418--432.

\end{thebibliography}

\end{document}